\documentstyle[aps,epsf,pre,floats,epsfig]{revtex}
\begin{document}
\title{Thin Film Diblock Copolymers in Electric Field:
Transition from Perpendicular to Parallel Lamellae}
\author{Yoav Tsori and David Andelman\\
School of Physics and Astronomy\\
Raymond and Beverly Sackler Faculty of Exact Sciences\\
Tel Aviv University, 69978 Ramat Aviv, Israel}
\date{24/4/2002}

\maketitle

\begin{abstract}
\setlength{\baselineskip}{24pt}

We examine the alignment of thin film diblock copolymers
subject to a perpendicular electric field. Two regimes are
considered separately: weak segregation and strong
segregation. For weakly segregated blocks and below a
critical value of the field, $E_c$, surface interactions
stabilize stacking of lamellae in a direction parallel to
the surfaces. Above the critical field, a first-order
phase transition occurs when lamellae in a direction
perpendicular to the confining surfaces (and parallel to
the field) become stable. The film morphology is then a
superposition of parallel and perpendicular lamellae. In
contrast to Helfrich-Hurault instability for smectic
liquid crystals, the mode that gets critical first has the
natural lamellar periodicity. In addition, undulations of
adjacent inter-material dividing surfaces are out-of-phase
with each other. For diblock copolymers in the strong
segregation regime, we find two critical fields $E_1$ and
$E_2>E_1$. As the field is increased from zero above
$E_1$, the region in the middle of the film develops an
orientation perpendicular to the walls, while the surface
regions still have parallel lamellae. When the field is
increased above $E_2$ the perpendicular alignment spans
the whole film. In another range of parameters, the
transition from parallel to perpendicular orientation is
direct.

\end{abstract}



\section{Introduction}\label{7:intro}
\setlength{\baselineskip}{24pt}

Diblock copolymers are known to self-assemble into a variety of
ordered structures, with a length scale ranging from nanometers to
micrometers
These phases have potential applications in nanolithographic
templates,$^{\cite{7:chaikin1}}$ waveguides $^{\cite{7:fink1}}$
and dielectric mirrors.$^{\cite{7:fink2}}$ The length scale and
morphology in the melt can easily be adjusted by controlling the
fraction $f=N_A/N$ of A monomers in an A/B chain of $N=N_A+N_B$
monomers, and the temperature
$T$.$^{\cite{7:Leibler80,7:O-K86,7:B-F90}}$

In experiments one often encounters samples in the lamellar phase
(made up of alternating planar A- and B-rich domains) in which the
melt is only partially ordered. Ordered microdomains, or grains
(typical size in the micrometer range), with grain boundaries
between them are defects that cost energy. In order to anneal
these defects, and to create a perfect alignment of the lamellae,
several techniques have been used. In the bulk, mechanical shear
proved to be a successful technique. Alignment by application of
an electric field
$^{\cite{7:amundson93,7:amundson94,7:russellSC96,7:TDR-MM00}}$
is also possible, but for macroscopic samples it requires high
voltage difference between the two bounding electrodes.
Nevertheless, this technique is especially suitable for thin
films, because the thickness involved makes the required large
fields (typically $10-30~{\rm V/\mu m}$) accessible.

We consider in this paper thin films of lamellar diblock
copolymers, under the influence of a perpendicular electric field.
Initially, the lamellae are parallel to the confining surfaces,
because of preferential short-range interactions with the
surfaces. In section \ref{7:wsl} we consider diblock copolymers in
the weak segregation regime. We show in section \ref{7:rowsm} that
electric field applied perpendicular to the surfaces can cause the
melt to transform from a parallel to a perpendicular orientation
through a first-order phase transition. The critical field $E_c$
for this transition is caused by a competition between the
electric field and surface interactions. In section \ref{7:ssl} we
investigate the thin-film alignment for diblocks in the strong
segregation regime. In this regime, the surface correlations are
finite, and thus the range of parallel ordering induced by the
surfaces is finite as well. We give the transitions between
parallel, perpendicular and mixed lamellae in terms of the system
parameters, using a phenomenological model. In both weak and
strong segregation regimes, large distortions are present in the
copolymer film, and these could be observed in experiments.

\section{Weakly segregated lamellae}\label{7:wsl}

The copolymer order parameter $\phi({\bf r})=\phi_A({\bf r})-f$ is
the deviation of the A monomer local volume fraction from its
average $f$. Above the order-disorder transition (ODT)
temperature, the melt is in the disordered, homogeneous state,
with $\phi({\bf r})=0$. As the temperature is reduced below the
ODT temperature, the system goes through a first order
phase-transition to the lamellar phase provided that $|f-\frac12|$
is small enough. Close to the ODT temperature and in the
single-mode approximation, the block copolymer (BCP) order
parameter is then given by
\begin{equation}\label{7:bulk}
\phi({\bf r})=\phi_L\cos({\bf
q_0\cdot r})
\end{equation}
where $d_0=2\pi/q_0$ is the period of lamellar
modulations, and $\phi_L$ is their amplitude (to be determined
later).

Consider a copolymer melt below the ODT confined by two
flat parallel surfaces at $y=\pm \frac12 L$, as in Fig.~1.
The surfaces reduce chain entropy, but also chemically
interact with the polymers. The difference in the A and
B-block surface interactions, $\sigma_{\rm AS}$ and
$\sigma_{\rm BS}$, defines the parameter $\sigma$,
\begin{equation}
\sigma=\sigma_{\rm AS}-\sigma_{\rm BS}
\end{equation}
In general, $\sigma(x,y)$ can be different for the two
surfaces, and hence $\sigma^+$ is defined on the
$y=\frac12 L$ surface and $\sigma^-$ is defined for
$y=-\frac12 L$. The surface interaction, $F_s$, can be
written as an integral over the bounding surfaces (in
units of $k_B T$)
\begin{eqnarray}\label{7:Fs}
{\cal F}_s=\int
\left\{\sigma^-\phi^-+\sigma^+\phi^+\right\}{\rm d}x{\rm
d}z
\end{eqnarray}
where $\phi^-=\phi(y=-\frac12 L)$ and $\phi^+=\phi(y=\frac12 L)$.
Terms which do not depend on the copolymer order parameter are not
important to subsequent calculations, and were dropped out. The
interaction in eq \ref{7:Fs} is short-range, localized at the
surfaces. A positive $\sigma^\pm>0$ induces adsorption of the B
monomers ($\phi^\pm<0$), and $\sigma^\pm<0$ the adsorption of the
A monomers ($\phi^\pm>0$) . We restrict ourselves to homogeneous
surfaces, for which $\sigma^\pm$ are taken to be constants over
each of the two surfaces.

First let us consider a BCP film without an external electric
field. We have already considered this case in a previous
publication,$^{\cite{7:epje01}}$ and briefly review here the BCP
behavior. If the surface affinities $\sigma^\pm$ are sufficiently
large, the lamellae will order in a parallel arrangement. These
lamellae stretch or compress, increasing the bulk free energy, in
order to decrease surface energy. We use below an adaptation of
the strong stretching approximation used by Turner
$^{\cite{7:turnerPRL92}}$ and Walton {\it et al.}
$^{\cite{7:W-RMM94}}$ to describe these lamellae. The lamellar
period is $d_0=2\pi/q_0$ and $m$ is the closest integer to
$L/d_0$. Depending on the values of $\sigma^\pm$, an integer
($n=m$) or half integer ($n=m+\frac12$) number of lamellae exist
between the two surfaces. In the former case the ordering is
symmetric (the same type of monomers wet both surfaces), while in
the latter it is antisymmetric (A monomers wet one surface,
whereas B monomers wet the other surface). The parallel lamellae
are described by an order parameter $\phi_\parallel$ given by
$^{\cite{7:epje01}}$
\begin{equation}\label{7:phipar}
\phi_\parallel(y)=\pm\phi_L\cos[q_\parallel(y+\frac12 L)]\\
\end{equation}
The wavenumber is $q_\parallel=2\pi n/L$, and the choice of $\pm$
sign in eq \ref{7:phipar} is such that the surface interactions,
eq \ref{7:Fs}, are minimized. The amplitude of sinusoidal
modulations $\phi_L$ is equal to the amplitude of density
modulations in a bulk system, and is given below [eq
\ref{7:phil}].

We consider now the case where an electric field is turned
on, in a direction that is perpendicular to the surfaces
($y$-axis in Fig.~1). Under conditions of constant voltage
difference across the electrodes situated at the two
bounding surfaces, the minimum of the free energy is
obtained by maximizing the capacitance. Noting that since
the A- and B-monomers (blocks) have different dielectric
constants, the effect of the electric field is to align
the BCP layers parallel to the field, i.e. perpendicular
to the surfaces. At a certain field strength, $E_c$, this
tendency balances the preference for parallel lamellae as
induced by the surfaces. Further increase of $E$ above the
critical value $E_c$ gives rise to a perpendicular
lamellar ordering.

Close to the ODT, the copolymer ordering is weak, and the
energetic cost of compressing or bending the lamellae is
small. In this regime the inter-material dividing surface
(IMDS) of the lamellar phase, given by the requirement
$\phi({\bf r})=\frac12-f$, can be substantially perturbed
from its flat state. This reasoning leads us to the
following superposition ansatz. For zero electric field
$E$, surface interactions orient the lamellae in a parallel
orientation. The order parameter is then given by
$\phi({\bf r}) =w(E=0)\phi_\parallel(y)$. The
dimensionless amplitude $w>0$ is determined by the
strength of the surface interactions,
 and can be larger than unity if
$\sigma^\pm$ are sufficiently large. Namely, the surface
induced order can be stronger than in the bulk. Upon
increase of the electric field, the function $w(E)$ of
this parallel state diminishes, while the function $g(E)$
of perpendicular lamellae (parallel to the electric field)
increases. This can be modeled by using the superposition
ansatz for the order parameter $\phi({\bf r},E)$ in the
presence of the field $E$:
\begin{equation}\label{7:ansatz}
\phi({\bf r},E)=w(E)\phi_\parallel({\bf r})+g(E)\phi_\perp({\bf r})
\end{equation}
The order parameter of the perpendicular lamellae is
$\phi_\perp(x)$. It depends only on the wavenumber $q_\perp$, and
is given in the single-mode approximation (weak segregation) by
\begin{equation}
\phi_\perp({\bf r})=\phi_\perp(x)=\phi_L\cos(q_\perp x)
\end{equation}
The wavenumber $q_\perp$ is yet to be determined. Without
the electric field, $E=0$, the amplitude of perpendicular
modulations vanishes, $g=0$. As the limit $E\to\infty$ is
approached, the effect of the confining surfaces becomes
negligible, and the BCP ordering is given by the bulk
perpendicular lamellae $\phi_\perp$ ($g=1$). Hence, the
weight amplitudes $w(E)$ and $g(E)$ satisfy the following
limits as function of $E$:
\begin{eqnarray}
g(E=0)=0,~~~~~g(E=\infty)=1\\
w(E=0)=const,~~~~~w(E=\infty)=0
\end{eqnarray}

In order to get explicit expression for the weight functions $w$
and $g$ we need to consider a specific model. The free energy we
use in this section is applicable to the weak segregation regime,
and is given by ${\cal F}={\cal F}_b+{\cal
F}_s$,$^{\cite{7:F-H87,7:epl01,7:mm01}}$ with ${\cal F}_s$ from eq
\ref{7:Fs}, and ${\cal F}_b$ being the bulk contribution. This
part of the free energy has a polymer (non-electrostatic) and
electrostatic contributions (in units of $k_B T$),
\begin{equation}
{\cal F}_b={\cal F}_p+{\cal F}_{\rm el}
\end{equation}
Hereafter we restrict ourselves to symmetric ($f=\frac12$) diblock
copolymers, where the polymer part of the free energy is
$^{\cite{7:Leibler80,7:F-H87,7:epl01,7:mm01,7:sh77,7:cc98}}$
%
\begin{eqnarray}\label{7:Fp}
{\cal F}_p= \int \left\{\frac12 \tau\phi^2+\frac12
h\left(q_0^2\phi
+\nabla^2\phi\right)^2+\frac{u}{24}\phi^4\right\}{\rm
d}^3r\label{7:Fpol}
\end{eqnarray}
The parameters in eq \ref{7:Fpol} are given by
\begin{eqnarray}
q_0&=& 1.95/R_g~~;~h=3\rho c^2 R_g^2/2q_0^2\\
\chi_c&\simeq&10.49/N~~;~\tau=2\rho N\left(\chi_c-\chi\right)
\end{eqnarray}
Denoting $b$ as the monomer size, the radius of gyration
for Gaussian chains is $R_g^2\simeq \frac16Nb^2$. The
polymerization index is $N$, the chain density of an
incompressible melt is $\rho=1/Nb^3$, and $\chi$ is the Flory
parameter. The amplitude
$\phi_L$ in the parallel and perpendicular states is
\begin{equation}\label{7:phil}
\phi_L^2=-8\tau/u~,~~~~~~\tau<0
\end{equation}
as is obtained by inserting eq \ref{7:bulk} in ${\cal F}_p$ of eq
\ref{7:Fp} and minimizing with respect to $\phi_L$. The
dimensionless ratio $u/\rho$ and $c$ are of order unity, and are
taken to be equal exactly to one in the remaining of the paper,
$u/\rho=c=1$. The correlation length $\xi\equiv
2q_0^2(|\tau|/h)^{-1/2}$ is assumed to be larger than the film
thickness, $\xi\gtrsim L$.$^{\cite{7:fraaije02}}$

The electrostatic contribution in units of $k_BT$ is
$^{\cite{7:amundson93,7:onuki95}}$
\begin{eqnarray}\label{7:Fel}
{\cal F}_{\rm el}=\beta\int (\hat{{\bf q}}\cdot{\bf
E})^2\phi_{\bf q}\phi_{-\bf q}{\rm
d}^3q\\
\beta=\frac{(\varepsilon_A-\varepsilon_B)^2}{4(2\pi)^4
k_BT\langle\varepsilon\rangle}\label{7:beta}
\end{eqnarray}
Here $\phi_{\bf q}$ is the Fourier transform of $\phi({\bf r})$:
$\phi({\bf r})=\int \phi_{\bf q} {\rm exp}(i{\bf q\cdot r}){\rm
d}{\bf q}$, and ${\bf \hat{q}}={\bf q}/q$ is a unit vector in the
${\bf q}$-direction. Copolymer modulations with a non-vanishing
component of the wavenumber ${\bf q}$ along the electric field,
have a positive contribution to the free energy. In other words,
there is a free energy penalty for having dielectric interfaces in
a direction perpendicular to the electric field. In eq
\ref{7:beta}, $\varepsilon_A$ and $\varepsilon_B$ are the
dielectric constants of the pure A and B-blocks, respectively,
$^{\cite{7:amundson93,7:amundson94}}$ and
$\langle\varepsilon\rangle$ is the material average dielectric
constant for the BCP film,
\begin{equation}
\langle\varepsilon\rangle=f\varepsilon_A+(1-f)\varepsilon_B
\end{equation}
Throughout the remaining of this paper we will focus only
on symmetric melts ($f=\frac12$), having an average
dielectric constant
$\langle\varepsilon\rangle=\frac12(\varepsilon_A+\varepsilon_B)$.
For small concentration variations, valid in the weak
segregation regime, $\varepsilon$ varies linearly with the
local copolymer composition $\phi$,
\begin{eqnarray}
\varepsilon(\phi)&=&(\frac12+\phi)\varepsilon_A+(\frac12-
\phi)\varepsilon_B
\nonumber\\&=&\langle\varepsilon\rangle+(\varepsilon_A-\varepsilon_B)
\phi
\end{eqnarray}

It is possible to perform the spatial integration in eqs.
\ref{7:Fs}, \ref{7:Fpol} and \ref{7:Fel}, yielding the free energy
per unit volume $F=F_p+F_{\rm el}+F_s$ for a general order
parameter $\phi({\bf r})$, eq \ref{7:ansatz}:
\begin{eqnarray}
F&=&\frac14\phi_L^2\left[\left(\tau+C_\parallel(E)\right)
w^2+\left(\tau+C_\perp\right) g^2\right]
\nonumber\\
&+&\frac{u\phi_L^4}{64}(w^4+g^4)+\frac{u\phi_L^4}{16}w^2g^2
+\frac{w\Sigma}{L}
\label{7:F}
\end{eqnarray}
The quantities $C_\parallel$ and $C_\perp$ are positive and
given by
\begin{eqnarray}
C_\parallel(E)&=&h(q_0^2-q_\parallel^2)^2+2\beta E^2\label{7:c_par}\\
C_\perp&=&h(q_0^2-q_\perp^2)^2\label{7:c_perp}
\end{eqnarray}
$\Sigma=\pm\phi_L\sigma^-\pm \phi_L\sigma^+$ is related to the
surface interaction, and is negative. The $\pm$ sign is determined
from the $\pm$ sign of the order parameter in eq \ref{7:phipar}.
The free energies $F_\parallel$  and $F_\perp$ of the parallel and
perpendicular states, respectively, are given as limiting cases
\begin{eqnarray}\label{7:f_par}
F_\parallel=F(w,g=0)&=&\frac14\phi_L^2\left[\tau+C_\parallel(E)\right]
w^2+\frac{u\phi_L^4}{64}w^4+\frac{w\Sigma}{L}\\
F_\perp=F(w=0,g=1)&=&\frac14\phi_L^2(\tau+C_\perp)
+\frac{u\phi_L^4}{64}\label{7:f_perp}
\end{eqnarray}

As discussed in the introduction, we concentrate on the
interesting case where in the absence of electric field, the BCP
has a parallel ordering given by $\phi=w\phi_\parallel$ (with some
weight $w$). This is equivalent to saying that there exist $w$
such that $F_\parallel(E=0)<F_\perp(E=0)$. By inserting $E=0$ and
$q_\perp=q_0$ in eqs. \ref{7:f_par} and \ref{7:f_perp} we get that
in this case
\begin{eqnarray}
\frac14\phi_L^2\left[\tau+C_\parallel(0)\right]w^2+\frac{u\phi_L^4}{64
}w^4
+\frac{w\Sigma}{L}~<~-\frac{\tau^2}{u}&&
\end{eqnarray}
Therefore, the free energy of the parallel lamellae is
expected to be lower than the free energy of the
perpendicular lamellae. This assumption is valid for
strong enough surface interactions, $\Sigma$. In the next
section we proceed to find the weight functions $w(E)$ and
$g(E)$ for any $E>0$.

\section{Results of the weak segregation
model}\label{7:rowsm}

The free energy (eq \ref{7:F}) is minimized with respect to $w$
and $g$ to yield two coupled algebraic equations:
\begin{eqnarray}\label{7:gov_g}
(\tau+C_\perp)g-\tau g^3-2\tau w^2g=0\\  \label{7:gov_f}
\left[\tau+C_\parallel(E)\right]w-\tau w^3-2\tau
g^2w+2\Sigma/L\phi_L^2=0
\end{eqnarray}
As we will see below, there is a critical value of the
electric field, $E_c$, which separates between the small
and large field behavior. One can estimate $E_c$ as the
field in which the bulk electrostatic energy $\sim LE^2$
(favoring perpendicular lamellae) balances the surface
interactions $\Sigma$, and hence $E_c\sim L^{-1/2}$.

Equations \ref{7:gov_g} and \ref{7:gov_f} are analyzed separately
for small and large
electric fields.\\ ~\\

\noindent {\bf (i) Small electric fields: $E<E_c$.}

For zero electric field, the lamellae are in their parallel state,
namely $g=0$. The solution with $g(E)=0$ and $w(E)\neq 0$ [in eqs.
\ref{7:gov_g} and \ref{7:gov_f}] corresponds to the minimum of the
free energy if $E$ is below a certain threshold value $E_c$, which
is to be determined later. Denoting $g_<(E)$ and $w_<(E)$ as the
weight amplitudes for electric fields $E<E_c$, they satisfy the
following equations:
\begin{eqnarray}
g_<(E)&=& 0\\
\left[\tau+C_\parallel(E)\right]w_<(E)-\tau
w_<^3(E)+2\Sigma/L\phi_L^2&=&0 \label{7:wg_below}
\end{eqnarray}

\noindent{\bf (ii) Large electric fields: $E>E_c$.}

There is a solution to eqs. \ref{7:gov_g} and \ref{7:gov_f} with
perpendicular lamellae,  namely with a nonzero $g(E)$. This
solution gives the minimum of the free energy above a critical
field, $E>E_c$. The weight amplitudes $g_>(E)$ and $w_>(E)$ are
then given by:
\begin{eqnarray}
g_>^2(E)~=~\frac{\tau+C_\perp}{\tau}-2w_>^2(E)~\geq~ 0\\
 \label{7:wg_above}
\left[-\tau+C_\parallel(E)-2C_\perp\right]w_>(E)+3\tau
w_>^3(E)+2\Sigma/L\phi_L^2~=~0
\end{eqnarray}
The above solution is valid provided that $w_>$ is small enough,
as given by the inequality
\begin{equation}\label{7:fc}
w_>^2(E)\leq\frac12+\frac{C_\perp}{2\tau}\leq\frac12
\end{equation}

Since $w(E)$ is a decreasing function, from eq \ref{7:c_perp} we
see that increasing the electric field $E$ from zero, the natural
mode of the perpendicular state, $q_\perp=q_0$, is the first to
become critical. Namely, the inequality above is obeyed first when
$q_\perp$ is equal to the bulk mode $q_0$, and only later by other
$q$-modes. For this reason we assume hereafter that $q_\perp
=q_0$, yielding $C_\perp=0$ and $g_>^2(E)=1-2w_>^2(E)$. Thus, BCP
modulations in a direction parallel to the surfaces have the bulk
(free) periodicity $d_0$.

Equation \ref{7:wg_above} is a cubic equation for $w_>$ and has an
analytical solution. It is convenient to express the solution via
a parameter $\theta$ defined as
\begin{equation}
\cos\left[\theta(E)\right]\equiv -\frac{\Sigma/L\phi_L^2}{3\tau}
\left(\frac{\tau-C_\parallel(E)}{9\tau}\right)^{-3/2}<0\label{7:theta}
\end{equation}
and take it to be in the range $\pi<\theta<\frac32\pi$. Recalling
that $\Sigma $ and $\tau$ are negative, the solution $w_>$ to eq
\ref{7:wg_above} is then simply given by
\begin{equation}
w_>(E)=2\sqrt{\frac{\tau-C_\parallel(E)}{9\tau}}\cos
\left(\frac{\theta}{3}\right)
\end{equation}

Curves of $g(E)$ and $w(E)$ are shown in Fig.~2 within the
assumption $q_\perp=q_0$. For zero electric field, the film has
parallel lamellae, $g(0)=0$. With increasing electric field $E$,
the amplitude $w_<(E)$ of parallel lamellae decreases
monotonically, as is given by eq \ref{7:wg_below}, while $g_<(E)$
remains zero. At the critical field, $E=E_c$, there is a
first-order phase transition, and perpendicular lamellae appear.
The weight amplitudes $w(E)$ and $g(E)$ are discontinuous, the
jump in their values is determined by the degree of segregation
$N\chi$, inter-surface separation $L$ and surface parameters
$\sigma^\pm$. Further increase of the electric field causes $g(E)$
to increase while $w(E)$ decreases. For very large electric
fields, $E\to\infty$, the perpendicular state saturates to its
value $g=1$, and parallel lamellae completely disappear, $w=0$.

The critical field $E_c$ is determined by the condition
\begin{equation}
F(w_<,g_<)=F(w_>,g_>)
\end{equation}
where the free energy $F$ is taken from eq \ref{7:F}. Namely, it
is the field where the two values of the free energy cross. The
critical field $E_c$ as a function of surface separation $L$ is
shown in Fig.~3 a for weak segregations ($N\chi=11$). As the
surface separation $L$ increases $E_c$ decreases with typical
oscillations of period $d_0$. These oscillations are caused by the
frustration occurring when the surface separation is
incommensurate with the lamellar period. Figure~3 b shows a
log-log plot of the same curve. The dashed line shows a fit to a
$E_c\sim L^{-1/2}$ scaling. This scaling can be obtained by
balancing the surface energy $\Sigma$ with the electrostatic
contribution $\sim LE_c^2$. There is good fit between this
$L^{-1/2}$ line and the peak positions, as expected for unstrained
films.$^{\cite{7:PW-MM99}}$ However, for frustrated films (where
the film thickness is incommensurate with the lamellar period) the
deviation from $E_c\sim L^{-1/2}$ becomes increasingly important
as the surface separation $L$ is reduced below roughly $6d_0$ for
the parameters used.

We define $w_<^c$ as the value of $w$ just before the
transition ($E\uparrow E_c$), and $w_>^c$ the value after
the transition ($E\downarrow E_c$),
and similarly for $g$. The jump in the weight amplitudes
$\Delta w$ and $\Delta g$ is
\begin{eqnarray}
\Delta w&=&|w_>^c-w_<^c|\\
\Delta g&=&g_>^c-g_<^c=g_>^c
\end{eqnarray}
These quantities are plotted as a function of surface separation
$L$ in Fig.~3 c, for the same parameters as in
Fig.~3 a. As $L$ increases, the critical field $E_c$
decreases, the jump in $w$ and $g$ gets larger, and the
transition from parallel to perpendicular lamellae becomes more
abrupt.

In Fig.~4 we show how the film changes its orientation and
morphology as the electric field increases. In part a we
show a contour plot of the copolymer order parameter
$\phi=w(E)\phi_\parallel+g(E)\phi_\perp$, for $E<E_c$, but
only slightly below it. The ordering in the film is
parallel to the surfaces, as $g(E)\equiv 0$. In part b the
field is slightly increased above its threshold value
$E_c$, and the undulations created by the appearance of the
perpendicular state are prominent. In contrast to the
classical Helfrich-Hurault undulations $^{\cite{7:hh}}$
calculated for smectic and cholesteric liquid crystals,
here the lateral wavelength is finite, and is equal to the
free periodicity $d_0$. Note that the correlation length
$\xi$ is larger than the film thickness, $\xi> L$. In
addition, the modulations of the adjacent inter-material
dividing surfaces (IMDS), given by $\phi({\bf r})=0$, are
shown in part c to be out-of-phase with each other and are
pronounced only for $E\simeq E_c$ . Although the
perpendicular ordering may be very strong, some parallel
ordering is still present (finite $w>0$). As the electric
field is further increased, there is only little
reminiscence of the parallel ordering, and the
perpendicular lamellae are nearly perfect. This can be
seen in Fig.~4 d, where we choose $E=4E_c$.

\section{Strongly segregated lamellae}\label{7:ssl}

In this section we consider the same alignment phenomenon
under an electric field as in the previous section, but
the BCP melt is assumed to be in the strong segregation
limit, i.e. $N\chi\gg N\chi_c\simeq 10.5$. In this regime
the lamellae are not easily deformed, and the effect of the
surface field is important only close to the confining
walls, in contrast to the long-range ordering induced by
the surfaces in the weak segregation.

\subsection{Unstrained films}\label{7:unstrained}

We discuss first the alignment phenomenon ignoring the
effect of incommensurability between the film thickness
$L$ and the free lamellar period $d_0$. In the following
section generalization to strained films will be presented
as well. For simplicity, we also assume that the two walls
are chemically identical, and preferring the B monomers:
$\sigma_{\rm BS}<\sigma_{\rm AS}$, $\sigma^\pm>0$. For
sufficiently strong $E$ fields we expect to nucleate a
region in the middle of the film with lamellae
perpendicular to the walls, as is shown schematically in
Fig.~5. Hence two regions of a ``T-junction'' morphology
will exist in the film, in the vicinity of the bounding
surfaces. A positive energy penalty (per unit area)
$\gamma_{_T}$ is associated with each of the two
T-junction defects. In principle, other types of defects
might exist in the film, but this can only affect the
value of $\gamma_{_T}$ and not the system behavior as is
described below.

We denote the size of the region that is perpendicular to the
surfaces by $l$. If $l=0$ the film has only parallel ordering,
while for $l=L$ the perpendicular ordering spans the whole film.
The free energies per unit area, $F_\parallel$ and $F_\perp$, for
these two extreme cases are:
\begin{eqnarray}\label{7:f_para_strong}
F_\parallel&=&LF_p+2\sigma_{\rm BS} -\lambda_\parallel
LE^2\\\label{7:f_perp_strong} F_\perp&=&LF_p+\sigma_{\rm
AS}+\sigma_{\rm BS}-\lambda_\perp LE^2
\end{eqnarray}
$F_p$ is the polymer free energy per unit volume of a BCP
in the lamellar phase, and in the above we have used two
different depolarization factors, $\lambda_\parallel$ and
$\lambda_\perp$, for the two orientations
\begin{eqnarray}
\lambda_\parallel&=&\frac{1}{4\pi}\frac{\varepsilon_A\varepsilon_B}
{\varepsilon_A+\varepsilon_B}\nonumber\\
\lambda_\perp&=&\frac{1}{16\pi}(\varepsilon_A+
\varepsilon_B)
\end{eqnarray}
These two depolarization factors can be obtained by
calculating the electrostatic energy $-(8\pi)^{-1}\int
\varepsilon E^2 {\rm d}^3r$ of the film and using the
boundary condition that the displacement field
$D=\varepsilon E$ is continuous across the two dielectric
boundaries. Note that $\lambda_\perp>\lambda_\parallel$,
meaning that $-\lambda_\perp E^2<-\lambda_\parallel E^2$,
and the perpendicular state is favored for $E\to\infty$.

Let us focus on the mixed state as is illustrated in
Fig.~5. For intermediate values of $l$, $0<l<L$, there are
two surface regions of parallel orientation and a central
region of perpendicular orientations. The free energy of
this mixed state $F_M$ (per unit area) is:
\begin{equation}\label{7:f_mixed_strong}
F_M=LF_p+2\sigma_{\rm BS}+2\gamma_{_T}-\lambda_m(l) LE^2
\end{equation}
The system can be regarded as being composed of two
capacitors of parallel lamellae and one capacitor of
perpendicular lamellae connected in series, yielding the
constant $\lambda_m(l)$
\begin{equation}
\lambda_m(l)=\left[\frac{L-l}{L\lambda_\parallel}+
\frac{l}{L\lambda_\perp}\right]^{-1}
\end{equation}
We note that a similar expression for the electrostatic part of
the mixed state free energy was previously derived by Pereira and
Williams. $^{\cite{7:PW-MM99}}$

The value of the surface energies $\sigma_{\rm AS}$,
$\sigma_{\rm BS}$ and $\gamma_{_T}$ must depend on $l$. To
see this consider, for example, $L\gg d_0$ and $l=\frac12
L$ . In this case $\gamma_{_T}$ is some constant. But as
$l\to L$, the value of $\gamma_{_T}$ must approach zero,
because when $l=L$ the T-junction does not exist, and the
energy associated with it is zero. We denote by $a$ the
cutoff length, which is the characteristic width at which
$\gamma_{_T}$ goes to zero. From the same reason
$\gamma_{_T}$ must also tend to zero as $l\to 0$, with
another cutoff length. For simplicity we assume that this
cutoff length is also equal to $a$. Similarly, the surface
interaction energies $\sigma_{\rm AS}$ and $\sigma_{\rm
BS}$ tend to zero as $l\to L$, with the same cutoff
length. The qualitative forms of these parameters are
shown in Fig.~6 a. The smooth monotonic decay to zero at
$l=L$ and $l=0$ is only suggestive. $\gamma_{_T}(l)$ has
oscillations with period $\frac12d_0$.

In principle, one should minimize $F_M$ with respect to the size
of perpendicular domain $l$. This generalized (mixed) state would
correspond to the parallel state when the minimum is at $l=0$, and
to the perpendicular state when the minimum is at $l=L$. However,
we do not know from a molecular description how $\sigma_{\rm AS}$,
$\sigma_{\rm BS}$ and $\gamma_{_T}$ fall off to zero, and
therefore use the approximation shown in Fig.~6 b. $\sigma_{\rm
AS}(l)=\sigma_{\rm AS}^0$ and $\sigma_{\rm BS}(l)=\sigma_{\rm
BS}^0$ are constant for $l<L-a$ and zero for $l>L-a$.
$\gamma_{_T}(l)=\gamma_{_T}^0$ is constant in the range $a<l<L-a$,
and zero otherwise. To make the notation simpler we drop hereafter
the superscript zero of $\sigma_{\rm AS}$, $\sigma_{\rm BS}$ and
$\gamma_{_T}$. Note that the effect of incommensurability between
the surface spacing $L$ and the lamellar period $d_0$, appears as
undulations in the plot of $\gamma_{_T}(l)$ (Fig.~6 a), and these
undulations are neglected here and will be addressed below
separately.

With the above assumptions, the size $l$ of a perpendicular domain
that minimizes the mixed free energy $F_M$ is $l=L-2a$. The free
energy of the mixed configuration is thus taken from eq.
\ref{7:f_mixed_strong} with $\lambda_m$ given by
\begin{equation}
\lambda_m=\left[\frac{2a}{L\lambda_\parallel}+
\frac{L-2a}{L\lambda_\perp}\right]^{-1}
\end{equation}

Using the assumption that  the B-polymer is adsorbed at
the surfaces in the parallel state ($\sigma_{\rm
BS}<\sigma_{\rm AS}$), we plot  in Fig.~7 $F_\parallel$,
$F_\perp$ and $F_M$ as a function of electric field
strength $E$.  The dimensionless parameter $\delta$
measuring the difference in A and B-block surface
interactions is defined as:
\begin{equation}\label{7:delta}
\delta\equiv \frac{\sigma}{\gamma_{_T}}=\frac{\sigma_{\rm
AS}-\sigma_{\rm BS}}{\gamma_{_T}}
\end{equation}
Based on the value of $\delta$, we now discuss two cases:\\

\noindent {\bf (i) Strong surface fields: $\delta>\delta^*$}.

If $\delta$ is larger than
a threshold value $\delta^*$, given by
\begin{eqnarray}\label{7:deltastar}
\delta^*&\equiv& 2\frac{\lambda_\perp-\lambda_\parallel}
{\lambda_m-\lambda_\parallel}\\
&=&2+ \frac{4\lambda_\perp}{\lambda_\parallel}\frac{a}{L}+
{\cal O}\left(\frac {a^2}{L^2}\right)\nonumber
\end{eqnarray}
there are two distinct critical fields $E_1$ and $E_2>E_1$
given by
\begin{eqnarray}
E_1&=&\left[\frac{2\gamma_{_T}}{L(\lambda_m-\lambda_
\parallel)}\right]^{1/2}
\label{7:e1a}\\\nonumber \\
E_2&=&\left[\frac{\gamma_{_T}(\delta-2)}{L(\lambda_\perp-
\lambda_m)}\right]^{1/2} \label{7:e2a}
\end{eqnarray}
The smallest of these two fields, $E_1$, obtained for
$F_M=F_\parallel$, (Fig. 7~a) is the field required to create the
T-junction defect in the film. This field is independent of
$\delta$, and for thick films ($L\gg a$) it scales as $E_1\sim
L^{-1/2}$. For $E$ fields in the range $E_1<E<E_2$, the film has a
region of size $l=L-2a$ with perpendicular lamellae, while
parallel lamellae are localized in a small region of size $a$ near
the two walls. The second field, $E_2$, obtained for
$F_M=F_\perp$, is larger than $E_1$ and corresponds to the
electric field that is required to destroy the parallel surface
layer (of width $a$). Note that although these fields are large
(typically between $1$--$30$ V/$\mu$m), they can be readily
achieved in thin-film experiments \cite{7:TDR-MM00}. For thick
films $E_2$ is independent of $L$ and obtains the asymptotic value
\begin{equation}
E_2\simeq
\left[\frac{\gamma_{_T}(\delta-2)\lambda_\parallel}
{2a(\lambda_\perp-
\lambda_\parallel)\lambda_\perp}\right]^{1/2}
\end{equation}

Note also the appearance of $\lambda_m-\lambda_\parallel$ and
$\lambda_\perp-\lambda_m$ in the denominator of the two critical
fields in eqs \ref{7:e1a} and \ref{7:e2a}. If the dielectric
contrast between A and B domains is small, namely
$\varepsilon_A/\varepsilon_B\approx 1$, then
$\lambda_\parallel\approx\lambda_\perp\approx\lambda_m$, and the
critical fields are large. If, on the other hand,
$\varepsilon_A/\varepsilon_B\gg 1$, then the critical fields
required to achieve mixed and perpendicular lamellae are small.
For many polymer surface combinations, $\gamma_{_T}$ is large and
therefore $\delta$ is small. In this case the system is classified
as having weak surface fields as is discussed below.
\\

\noindent {\bf (ii) Weak surface fields: $\delta<\delta^*$}.

For small values of $\delta$, namely $\delta<\delta^*$,
the free energies $F_\perp$ and $F_\parallel$ are always
smaller than $F_M$. Namely, the mixed state is not a
possible minimum of the film free energy. There is only
one transition at a critical field $E_3$ occurring when
$F_\perp$ intersects $F_\parallel$ (Fig.~7 b). This field
is given by:
\begin{eqnarray}\label{7:e1b}
E_3&=&\left[\frac{\gamma_{_T}\delta}{L(\lambda_\perp-\lambda_
\parallel)}\right]^{1/2}\nonumber\\
&=&\left[\frac{\sigma}{L(\lambda_\perp-\lambda_
\parallel)}\right]^{1/2}\sim~\delta^{1/2}~L^{-1/2}
\end{eqnarray}
When $E=E_3$, a direct transition occurs from a state
where the whole film has parallel lamellae ($E<E_3$) to a
state where the whole film has perpendicular lamellae
without any surface regions ($E>E_3$).

The system behavior can be summarized in a phase diagram which
depends on the three system parameters: the electric field $E$,
the surface interaction parameter $\delta$ and the film thickness
$L$. In Fig.~8 a, we show a 2-dimensional cut through the phase
diagram, varying $E$ and $\delta$ while keeping $L$ constant. For
small electric fields, the lamellae are in the fully parallel
configuration. If $\delta$ is small, a first-order phase
transition to the fully perpendicular state occurs when $E$ is
increased above $E_3$, eq~\ref{7:e1b}. If $\delta$ is large
enough, namely $\delta>\delta^*$, there are two transitions when
the field is increased. In the regime $E_2>E>E_1$, the film is in
a mixed state, and layers of thickness $a$ with parallel lamellae
still exist close to the surfaces. As $E$ is further increased
above $E_2$, the film has a fully perpendicular state. Note that
when $\delta>\delta^*$, $E_1$ is the field required to initiate
the T-junction defect, and so is independent of $\sigma$, while
$E_2$ is the field that destroys the surface layer, and therefore
$E_2\sim (\sigma-2\gamma_{_T})^{1/2}$.

As we have seen, the condition $\delta>\delta^*$ is
required for the existence of the mixed state, and hence
for the existence of two critical fields $E_1$ and $E_2$.
Alternatively, this condition can be viewed as a
restriction on the film thickness $L$ for a given $\delta$.
Therefore, a mixed state does not exist if
\begin{equation}
L<L^*\equiv
\frac{2a(\lambda_\perp-\lambda_\parallel)[2\lambda_\perp+
(\delta-2)\lambda_\parallel]}{\lambda_\parallel
[(\delta+2)\lambda_\perp +(\delta-2)\lambda_\parallel]}
\end{equation}
In the limit $\delta\gg 1$ and in the case of polystyrene
($\varepsilon\approx 2.5$) and polymethyl methacrylate
($\varepsilon\approx 6$) diblock copolymer, one obtains
$L^*=2a(\lambda_\perp-\lambda_\parallel)/(\lambda_\parallel
+\lambda_\perp)\approx 1.3 a$. It is smaller for smaller values of
$\delta$. Hence the mixed state exists for film thickness larger
than the range of surface induced ordering $a$. We see here again
why this mixed state is not expected to occur in the weak
segregation limit.

An alternative cut through the phase diagram is shown in Fig.~8 b,
where $\delta$ is fixed while $E$ and $L$ are allowed to vary. For
small surface separations, $L<L^*$, there is a transition from
parallel to perpendicular lamellae at $E=E_3$ [eq \ref{7:e1b}].
For larger separations, $L>L^*$, the mixed state appears at
$E=E_1$ [eq \ref{7:e1a}]. The transition to a fully perpendicular
state occurs at $E=E_2$ [eq \ref{7:e2a}]. The main difference
between this diagram and the one obtained by Pereira and
Williams,$^{\cite{7:PW-MM99}}$ is that we always find the
perpendicular state to be favored for large enough electric
fields.

\subsection{Strained films}

Strained films occur when the film thickness $L$ does not match
the lamellar period $d_0$, namely $L/d_0$ is not an integer or a
half-integer number. The effect of this mismatch, which was
neglected in the previous section, is considered below. The strong
segregation theory of Turner $^{\cite{7:turnerPRL92}}$ and Walton
{\it et al} $^{\cite{7:W-RMM94}}$ has been successful in
describing confined lamellae, and therefore we use it in this
section to include the effect of lamellae frustration, and to
modify the free energy $F_\parallel$ from section
\ref{7:unstrained}.
We use the convention that $m$ is the closest integer to $L/d_0$,
yielding $F_\parallel$ which is the minimum of $F^s_\parallel$ and
$F^{\rm as}_\parallel$, the symmetric and antisymmetric parallel
states free energies, respectively:
\begin{eqnarray}
F^s_\parallel &=&\sigma_{\rm
AB}\left[\left(\frac{L}{d_0}\right)^3\cdot \frac{1}{n^2}+2n\right]
+2\sigma_{\rm BS}-\lambda_\parallel LE^2 \nonumber\\
&& {\rm for}~~ n=m,~~{\rm symmetric}\label{7:48}\\ && \nonumber\\
F^{\rm as}_\parallel &=&\sigma_{\rm
AB}\left[\left(\frac{L}{d_0}\right)^3\cdot \frac{1}{n^2}+2n\right]
+\sigma_{\rm AS}+\sigma_{\rm BS}-\lambda_\parallel LE^2 \nonumber
\\&& {\rm for}~~ n=m\pm \frac12,~~{\rm antisymmetric}\label{7:49}
\end{eqnarray}
Note that in the above expressions for the free energy, eqs
\ref{7:48} and \ref{7:49}, the electrostatic part is exactly like
in Sec. \ref{7:unstrained}. In the symmetric state, both surfaces
are wetted by B monomers, while in the antisymmetric state A
monomers adsorb to one surface and B monomers adsorb to the second
surface.

In Fig.~9 we present the phase diagrams when the mismatch between
$L$ and $d_0$ is taken into account, plotted with same parameters
as in Fig.~8. For very thin films, $L<d_0$, the perpendicular
state is favored over the parallel state, even for $E=0$. Apart
from this, Fig.~9 a is similar to Fig.~8~a. However, Fig.~9 b is
different than Fig.~8~b. The critical fields $E_1$ and $E_3$
separating the parallel state from the other two states have
oscillations with period $d_0$. These curves are similar to the
critical field curve for weakly segregated lamellae, Fig.~3,
computed in Sec.~3, and also to the curve calculated by Ashok {\it
et al.} using a similar model and different
parameters.$^{\cite{7:ashok}}$ Note that from the discussion of
unstrained films in section \ref{7:unstrained}, $E_1$ and $E_3$
have oscillations around lines which decay as $ L^{-1/2}$ for
large $L$. The field $E_2$ separating the mixed state with the
perpendicular state depends only weakly on $L$.

\section{Conclusions}\label{7:conclusions}

In this paper we study the influence of an applied
electric field on the morphology of thin film diblock
copolymers. In the absence of an electric field the
lamellae are taken to be ordered parallel to the confining
surfaces. Strong enough $E$ fields in the direction
perpendicular to the surfaces will eventually orient the
lamellae in a perpendicular direction. However, the
response of weakly segregated lamellae is different than
the response of strongly segregated lamellae. In the
former case, the BCP order parameter is obtained as a
function of electric field strength $E$ and other system
parameters (such as the degree of segregation $N\chi$,
strength of surface interactions $\sigma^\pm$ and film
thickness $L$). The field applied in the perpendicular
direction diminishes the amplitude of the parallel BCP
state. Above the critical field, $E>E_c$, a first-order
phase transition occurs, from the parallel into the
perpendicular state.

The first $q$-mode which becomes stable in the perpendicular state
$q_\perp$ has a finite periodicity, equal to the bulk spacing
$d_0=2\pi/q_0$. Moreover, modulations of adjacent inter-material
dividing surfaces (IMDS) are out-of-phase with each other.  This
is different than the strong segregation instability studied by
Onuki and Fukuda,$^{\cite{7:onuki95}}$ where the slowly varying
phase $\phi=\phi_L\cos[q_0x+u(y)]$ is used, and the free energy is
expanded in small $u$. At the onset of instability, they found
that adjacent IMDS lines are in-phase with each other. The weak
segregation IMDS undulations studied here appear because in this
regime it is relatively easy to deform the lamellae. For $E>E_c$,
parallel and perpendicular lamellae coexist in the film. This
superposed state is yet to be verified in experiments.

The critical field $E_c$ as a function of inter-surface
separation $L$ decays with characteristic oscillations of
period $d_0$. These oscillations are the result of
lamellar frustration, occurring when the period
$2\pi/q_\parallel$ is different than $d_0$ (while
$q_\perp$ is taken always to be equal to $q_0$). The
deviation of the critical field $E_c$ from the $E_c\sim
L^{-1/2}$ scaling (of unstrained films) is important only
for small surface separations. For large $L$ values, the
$L^{-1/2}$ scaling describes well the system behavior.

The weak segregation treatment we present relies on mean-field
theory. It is valid close to the critical point, but not too close
where critical fluctuations become important.$^{\cite{7:braz}}$ We
have ignored the deviations of the inter-material dividing surface
from the perfect flat shape that occur near the surfaces. These
deviations can be important for small surface separations
($d_0\gtrsim L$), or close to the ODT,$^{\cite{7:epl01,7:mm01}}$
and should be properly accounted for.

In the second part of the paper we consider BCP film in the strong
segregation regime. In contrast to the weak segregation case, here
the surface induced ordering has a finite range of length $a$.
This added length (taken in the weak segregation limit to be much
larger than the film thickness $L$) results in the existence of
two critical fields. Provided that the parameter $\delta$ is large
enough, $\delta>\delta^*$ [eq \ref{7:deltastar}], for small
electric fields, $E<E_1$, the film has parallel lamellae. As the
field is increased above $E_1$ but below $E_2$, a region of size
$L-2a$ of perpendicular lamellae nucleates in the middle of the
film. In this mixed state, a layer of parallel lamellae of
thickness equal to the cutoff length $a$ still persists close to
the surfaces. When $E>E_2$ the system is in the perpendicular
state.

The full phase diagram in the $E$-$\delta$ and the $E$-$L$ planes
is given in Fig.~8. As discussed above, for large
$\delta>\delta^*$ there is a transition from parallel to mixed
lamellae at $E=E_1$ [eq \ref{7:e1a}] and from mixed to
perpendicular lamellae at $E=E_2$ [eq \ref{7:e2a}]. Another
scenario of a direct transition from a parallel to a perpendicular
state at $E=E_3$ [eq \ref{7:e1b}] is realized for small $\delta$,
$\delta<\delta^*$. These diagrams do not take into account the
frustration effects caused by a mismatch between the wall
separation and the natural lamellar periodicity in the bulk. In
this respect, they are qualitatively applicable to non-polymeric
systems such as ferrosmectics in magnetic
fields,$^{\cite{7:fabre}}$ and to BCP in the hexagonal phase, as
investigated experimentally. $^{\cite{7:TDR-MM00}}$ The difference
between the hexagonal and lamellar phase behavior comes from the
different value of the depolarization factors $\lambda_\parallel$,
$\lambda_\perp$, and different numerical coefficients in
eqs.~\ref{7:f_para_strong}, \ref{7:f_perp_strong} and
\ref{7:f_mixed_strong}. These modifications only change the
critical fields $E_1$, $E_2$ and $E_3$ by a numerical factor, but
do not alter the phase behavior. Indeed, two critical fields
between which a regime of mixed state exists have been found
experimentally for a BCP is its hexagonal phase by Russell and
co-workers.$^{\cite{7:TDR-MM00}}$  In their experiment, the
threshold field (called here $E_2$) for transition from the mixed
to the fully perpendicular state has no noticeable dependence on
surface separation $L$, in accord with our calculations as shown
in Fig.~8 b and Fig.~9 b.

When the surface separation $L$ is not an integer or half integer
number of the bulk lamellar period $d_0$, the phase diagrams are
given by Fig.~9. While in the $E$-$\delta$ plane the behavior is
only slightly changed, in the $E$-$L$ plane the border of the
parallel state has prominent oscillations with period $d_0$.
Finally, we have not considered the copolymer density undulations
as was done by Onuki and Fukuda.$^{\cite{7:onuki95}}$ These might
change the boundary lines between the parallel, perpendicular and
mixed phases.

The main prediction of this paper is the dependence of the
transition fields on the block copolymer film thickness $L$ and
the phase diagrams of Fig. 9. It will also be of interest to check
experimentally under what conditions the mixed state exists. In
addition, the energy penalty $\gamma_{_T}$ of the defect created
in the film as well as the difference in surface energies of the
two blocks, $\sigma=\sigma_{\rm AS}-\sigma_{\rm BS}$, can be
deduced in experiment from the measured values of $E_1$, $E_2$ and
$E_3$ using eqs. \ref{7:delta}, \ref{7:e1a} and \ref{7:e2a}.

\section*{Acknowledgments}
We would like to thank T. Taniguchi and T. Yukimura for
critical remarks. We benefitted from fruitful discussions
with B. Ashok, M. Cloitre, J. DeRouchey, M. Doi, L.
Leibler, M. Muthukumar, R. Rosensweig, T. P. Russell, M.
Schick, F. Tournilhac and T. Thurn-Albrecht. Partial
support from the U.S.-Israel Binational Foundation
(B.S.F.) under grant No. 98-00429 and the Israel Science
Foundation founded by the Israel Academy of Sciences and
Humanities
--- centers of Excellence Program is gratefully
acknowledged.



\newpage

\begin{itemize}

\item{\bf Figure~1:} Schematic illustration of the system. The two
confining surfaces are at $y=\pm \frac12 L$, and have surface
interaction parameters $\sigma^\pm$. The electric field points in
the perpendicular $y$ direction, and is produced by a potential
difference between the two surfaces (electrodes). \label{7:1}

\item{\bf Figure~2:} Weight functions $w(E)$ and $g(E)$.
The horizontal dash-dot line $w=1/\sqrt{2}$ is the value of $w$
below which a nonzero $g(E)$ is possible. Surface separation is
$L=8d_0$ and the Flory parameter is $N\chi=11$. The film is
symmetric, $\sigma^+=\sigma^-=0.6hq_0^3\phi_L$. \label{7:2}

\item{\bf Figure~3:} (a) Critical field $E_c$ [in units of
$(hq_0^4/\beta)^{1/2}$] required for the appearance of
perpendicular lamellae, as a function of surface separation $L$.
(b) log-log plot of the same curve as in part a. The dashed line
is a straight line corresponding to $E_c\approx 0.5(L/d_0)^{-1/2}$
(see text). (c) The jumps $\Delta g$ (dashed line) and $\Delta w$
(solid) in the amplitudes $g$ and $w$ at the critical field $E_c$,
as a function of surface separation $L$. The film is symmetric,
and $\sigma^\pm$ and $N\chi$ are as in Fig.~2. \label{7:3}

\item{\bf Figure~4:} Contour plots of the BCP order parameter
$\phi(x,y)=w(E)\phi_\parallel(y)+g(E)\phi_\perp(x)$ for symmetric
film. The surfaces are at $y=\pm \frac12 L=\pm 2d_0$, and the
field is in the $y$ direction. In part a the field is a little
smaller than the critical field, $E=0.98E_c$, and the film has a
perfect parallel ordering. In part b the field is just above the
threshold, $E=1.02E_c$. The film morphology is a superposition of
parallel and perpendicular lamellae. (c) A plot of the IMDS [given
by $\phi(x,y)=0$] of part b. In part d $E=4E_c$, and the lamellae
are in the perpendicular state with small distortions. The surface
fields are $\sigma^+=\sigma^- =0.5hq_0^3\phi_L$, and the Flory
parameter is $N\chi=11$ corresponding to correlation length
$\xi\simeq 4.8d_0>L$. The B monomers (colored black) are attracted
to the two symmetric surfaces. \label{7:4}

\item{\bf Figure~5:} Illustration of the lamellae in the film in the
strong segregation regime. A region with perpendicular lamellae
exists in the middle of the film, with size $l$. Two regions of
size $a$ each are the parallel lamellar regions. \label{7:5}

\item{\bf Figure~6:} (a) Qualitative dependence of $\gamma_{_T}$
and $\sigma_{\rm BS}$ on the thickness of perpendicular domain
$\ell$, defined in Fig.~5. (b) Simplified curves of part a. The
cutoff width is $a$, the total film thickness is $L$, and
$\gamma^0_{_T}$ and $\sigma^0_{BS}$ are the mean values of
$\gamma_{_T}$ and $\sigma_{BS}$, respectively, for $a<\ell<L-a$.
\label{7:6}

\item{\bf Figure~7:} Sketch of the dependence of $F_\parallel$,
$F_\perp$ and $F_M$ on the field $E$, in solid line, circles and
dashed line, respectively. (a) $\delta>\delta^*$: There are two
critical fields, obtained when $F_M(E_1)=F_\parallel(E_1)$ and
$F_M(E_2)=F_\perp(E_2)$. (b) $\delta<\delta^*$: $F_M(E)$ is always
larger than $F_\perp(E)$, and there is only one critical field
obtained when $F_\parallel(E_3)=F_\perp(E_3)$. Above this field
the most stable state is the perpendicular ordering. \label{7:7}

\item{\bf Figure~8:} (a) Phase diagram in the $E$-$\delta$ plane. If
$\delta=(\sigma_{\rm AS}-\sigma_{\rm BS})/\gamma_{_T}<\delta^*$,
there is a transition between parallel and perpendicular lamellae
at $E=E_3$. For $\delta>\delta^*$, there is a transition from the
parallel to the mixed state at $E=E_1$, followed by a second
transition from the mixed to the perpendicular state when
$E=E_2>E_1$. Surface separation is chosen as $L=10d_0$ and
$a=d_0$. (b) Similar diagram, but in the $E$-$L$ plane, with
$\delta=5$. In both parts, the effect of mismatch between $L$ and
$d_0$ is neglected, and the electric fields are scaled by
$(\gamma_{_T}/d_0)^{1/2}$. \label{7:8}

\item{\bf Figure~9:} Phase diagrams as in Fig.~8, but for
strained films. The critical fields $E_1$ and $E_3$ have
oscillations as a function of $L$, in contrast to Fig.~8 b, but
follow on average a similar $\sim L^{-1/2}$ scaling behavior.
\label{7:9}

\end{itemize}


\newpage

\begin{figure}[t]
\includegraphics[scale=0.5,bb=20 160 540 760,clip]{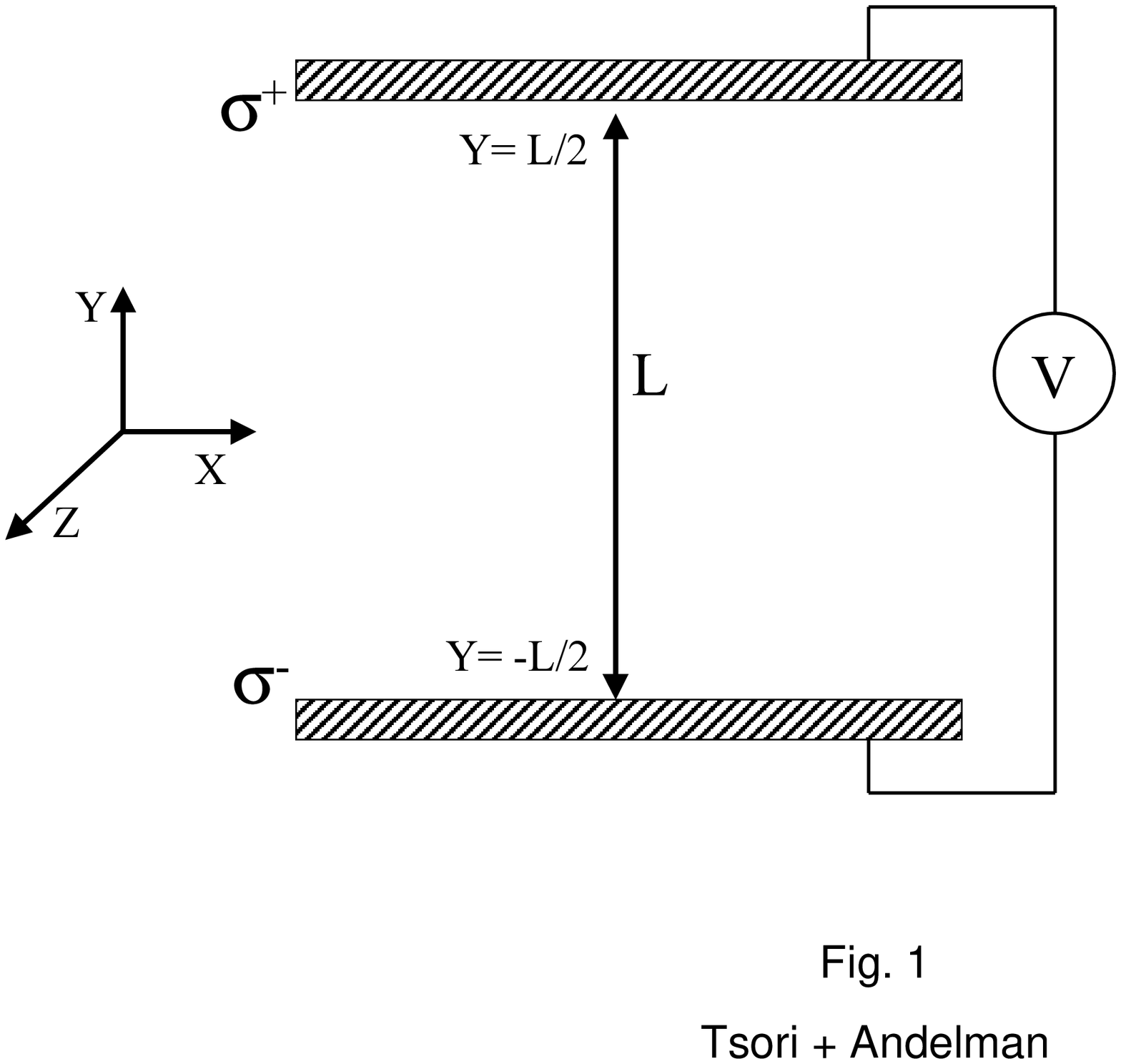}
\end{figure}

\begin{figure}[t]
\includegraphics[scale=0.65,bb=95 120 470 560]{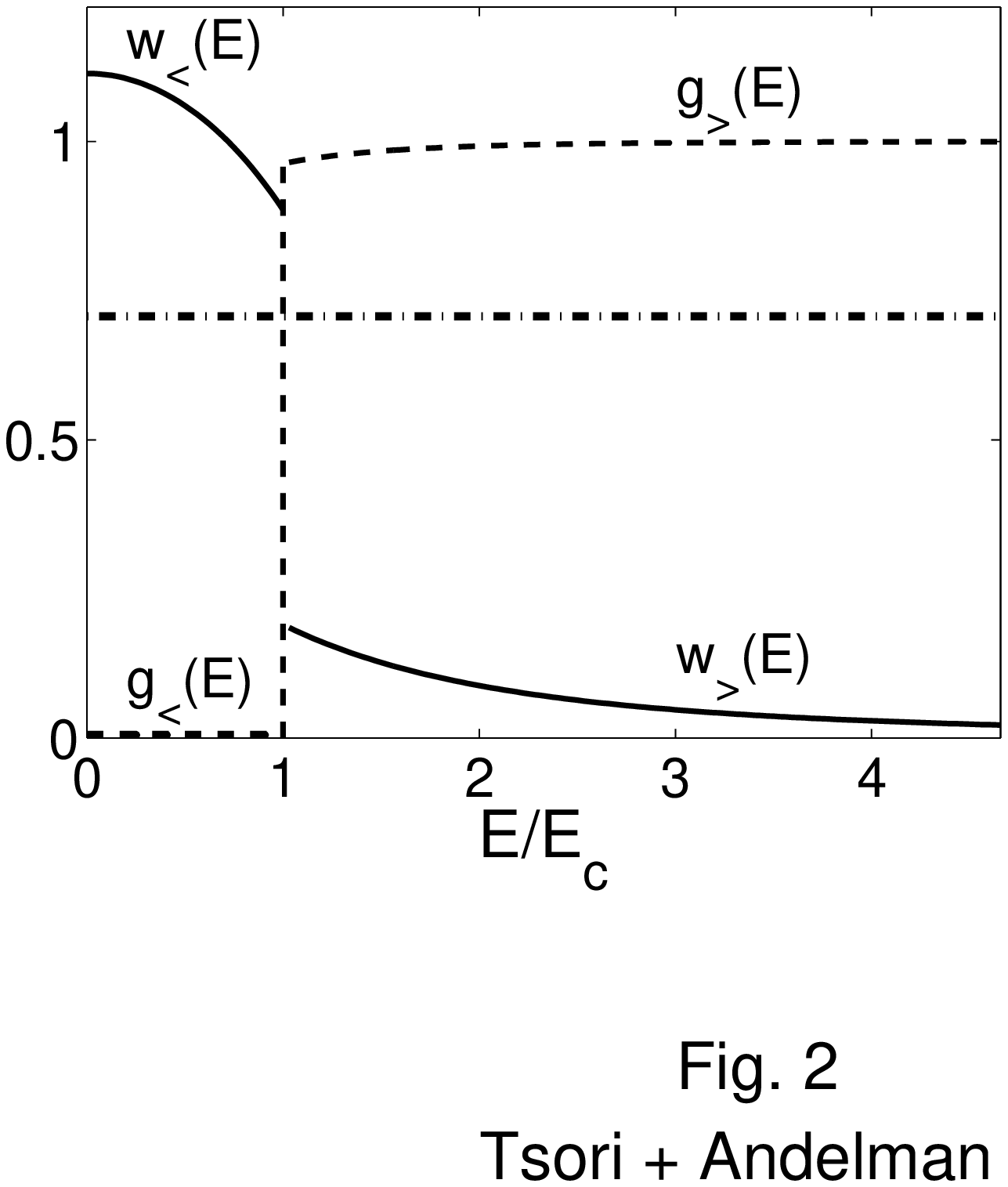}
\end{figure}

\begin{figure}[t]
\includegraphics[scale=0.8,bb=15 120 580 720,clip]{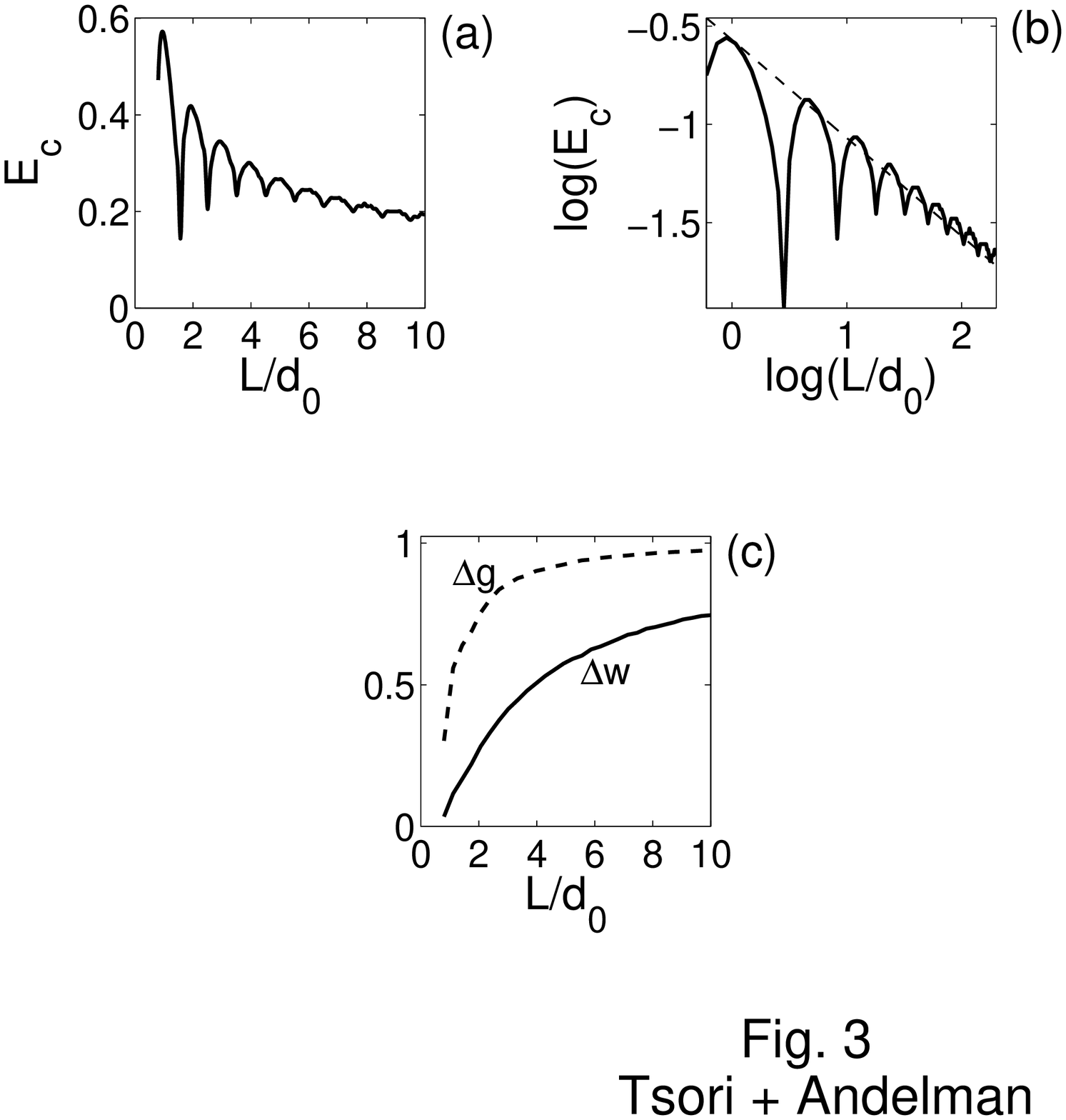}
\end{figure}

\begin{figure}[t]
\includegraphics[scale=0.75,bb=30 18 570 730,clip]{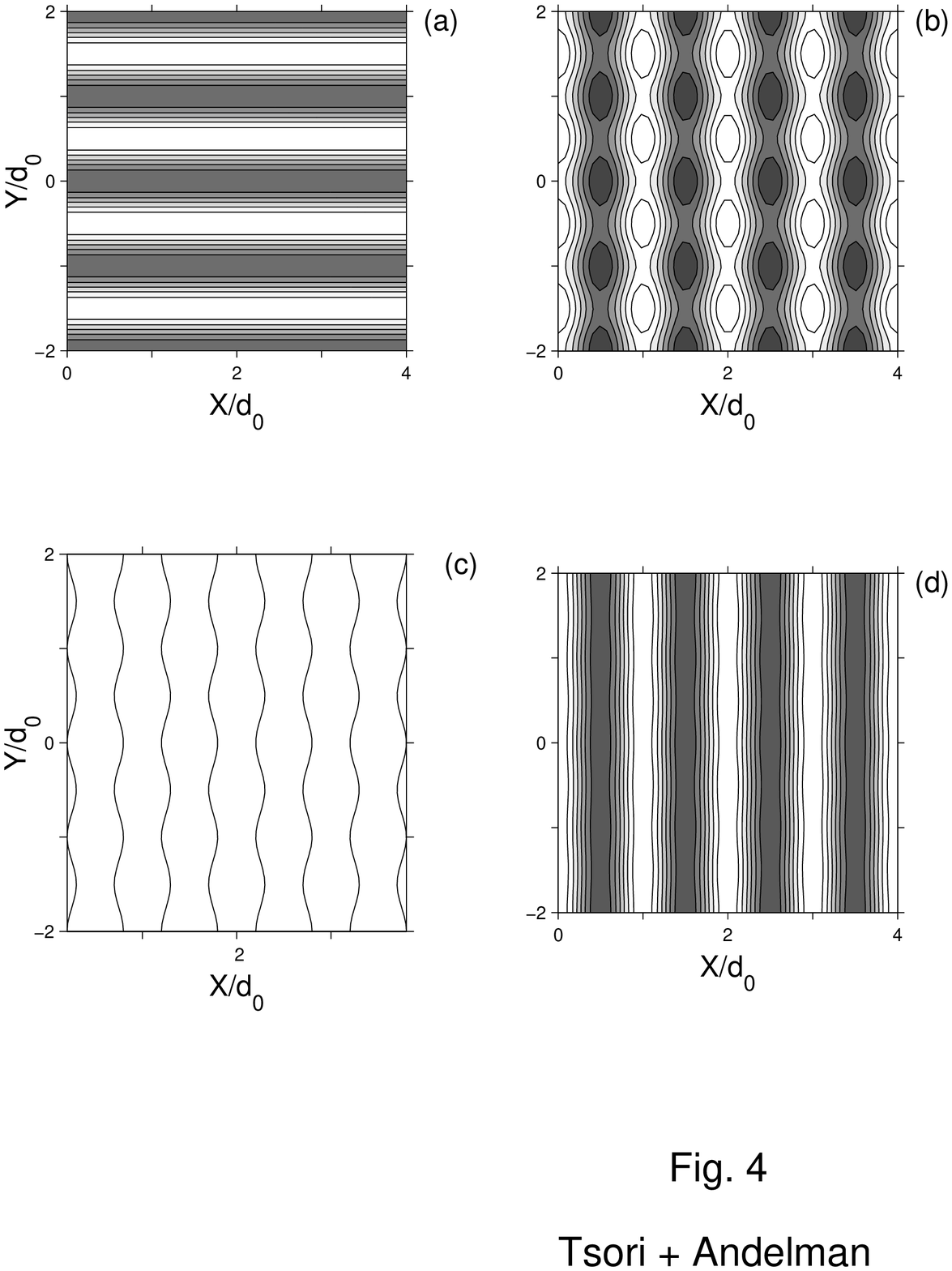}
\end{figure}

\begin{figure}[t]
\includegraphics[scale=0.5,bb=130 230 525 770,clip]
{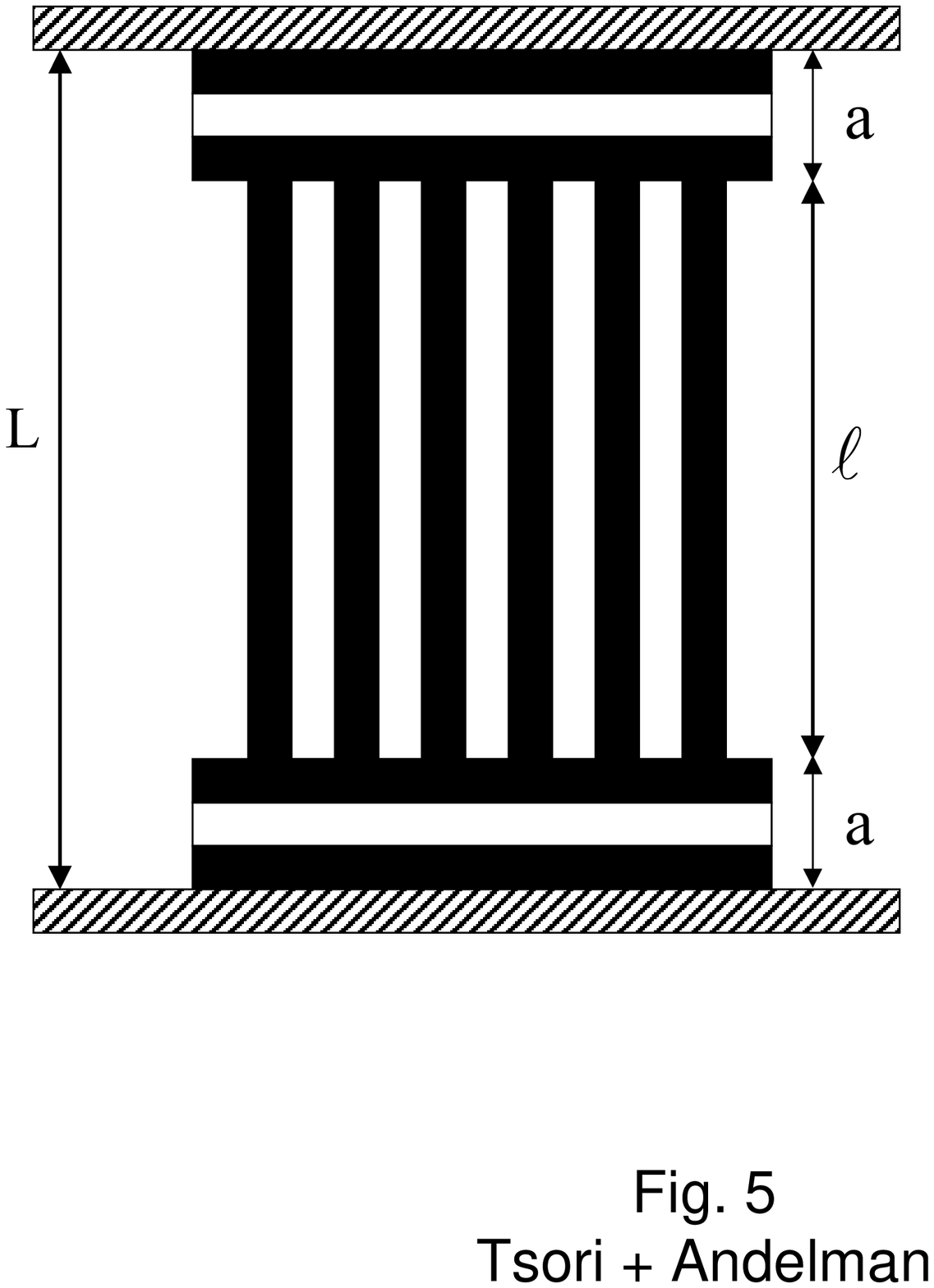}
\end{figure}

\begin{figure}[t]
\includegraphics[scale=0.8,bb=50 270 590 730,clip]
{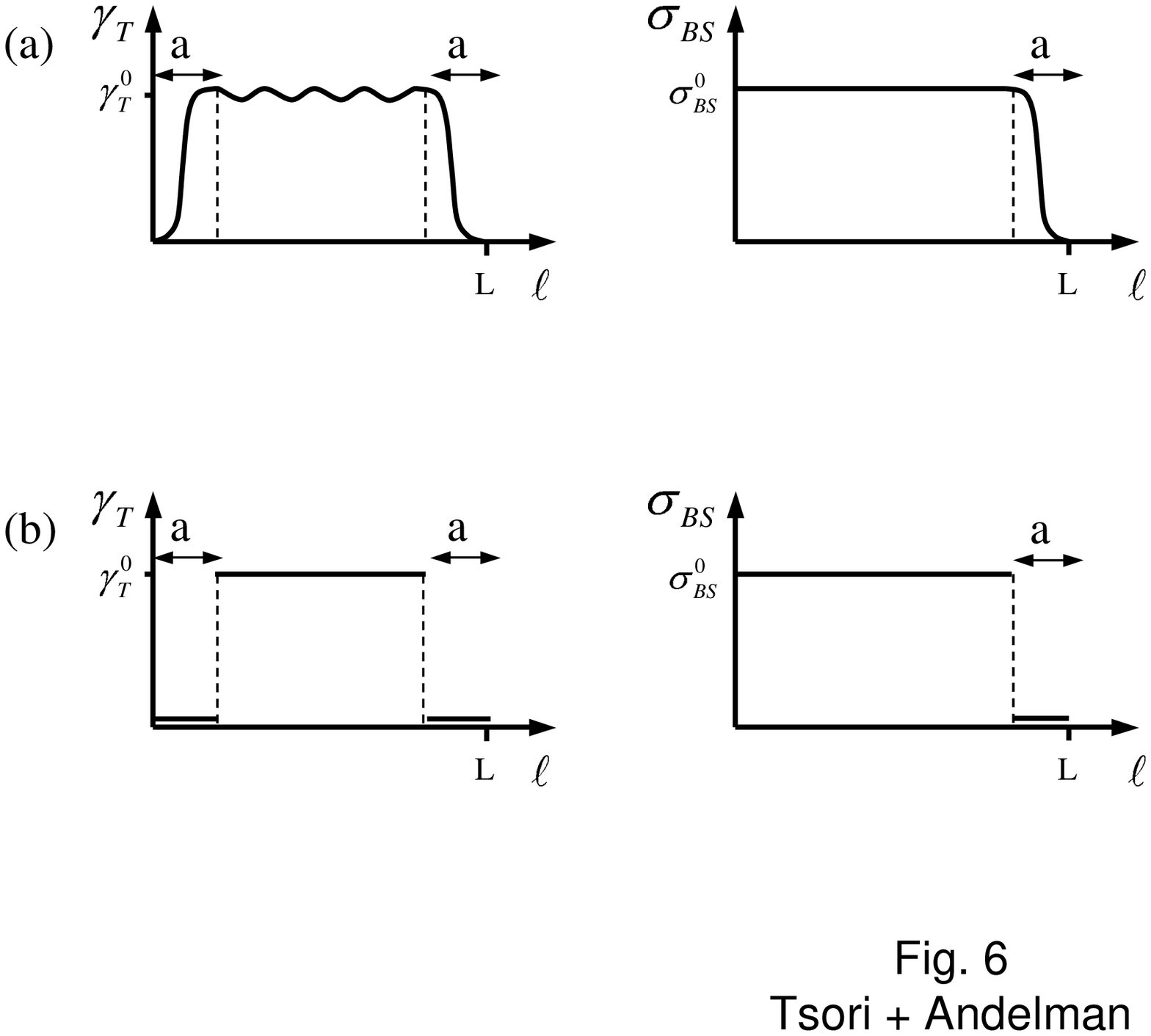}
\end{figure}

\begin{figure}[t]
\includegraphics[scale=0.55,bb=120 70 520 830,clip]{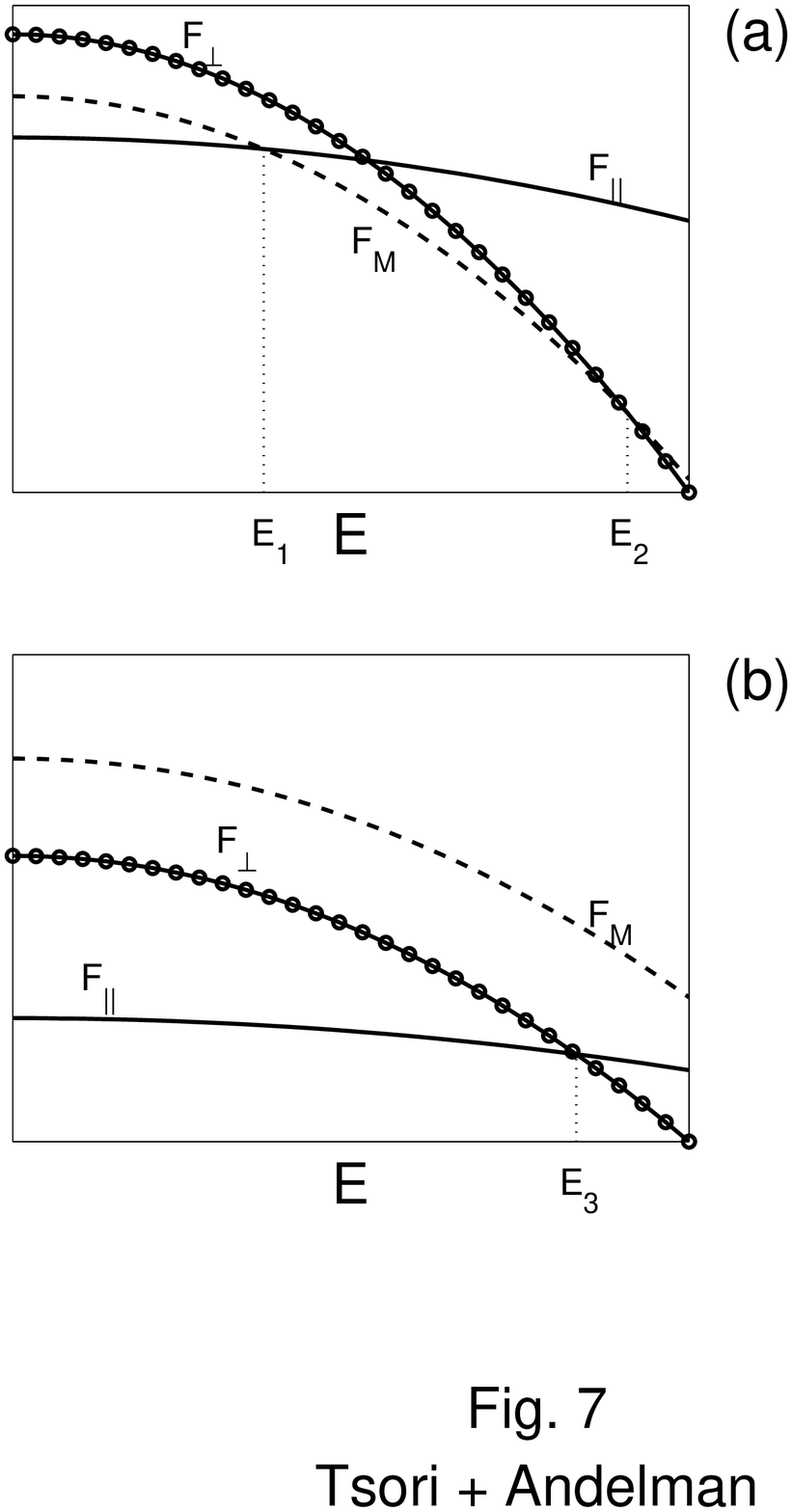}
\end{figure}

\begin{figure}[t]
\includegraphics[scale=0.9,bb=145 130 435 810,clip]{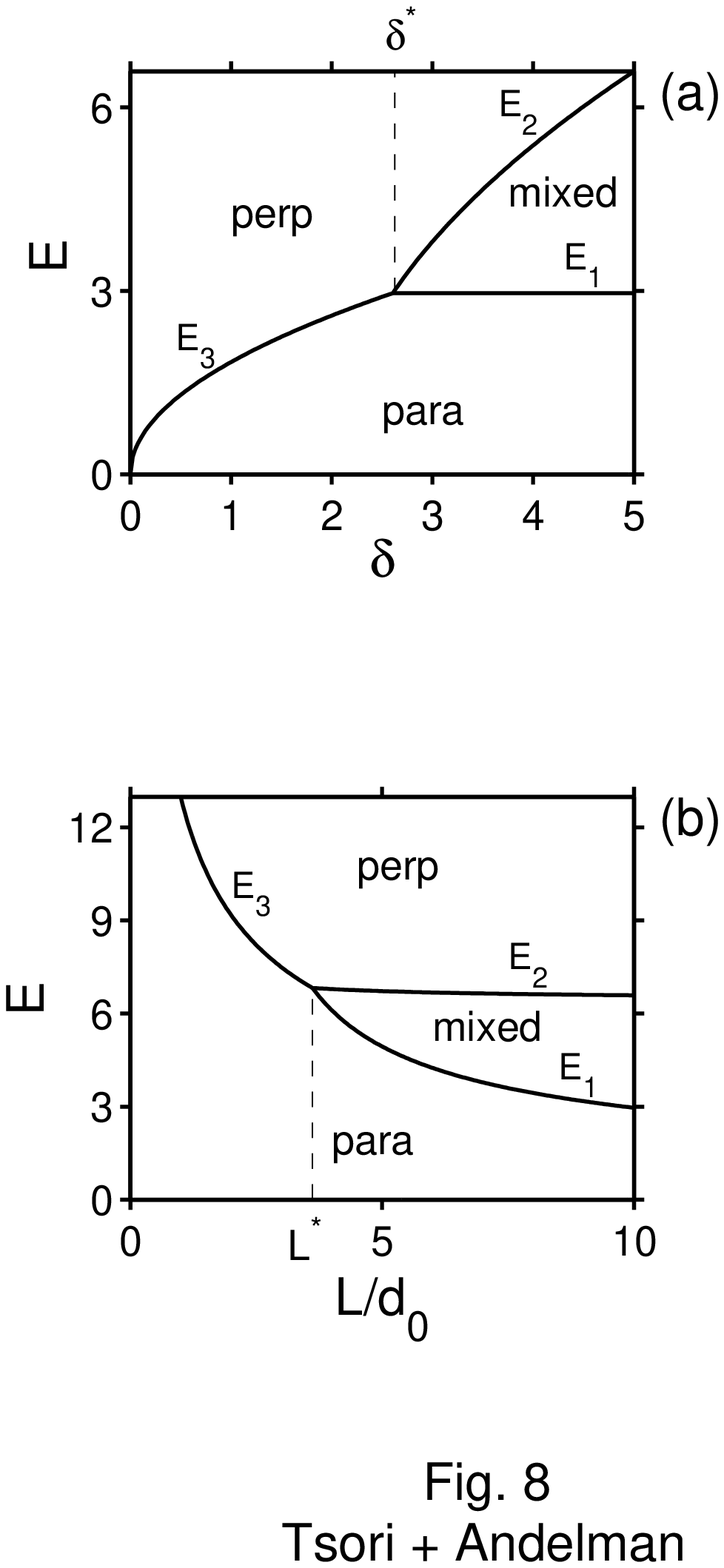}
\end{figure}

\begin{figure}[t]
\includegraphics[scale=0.8,bb=130 70 450 800,clip]{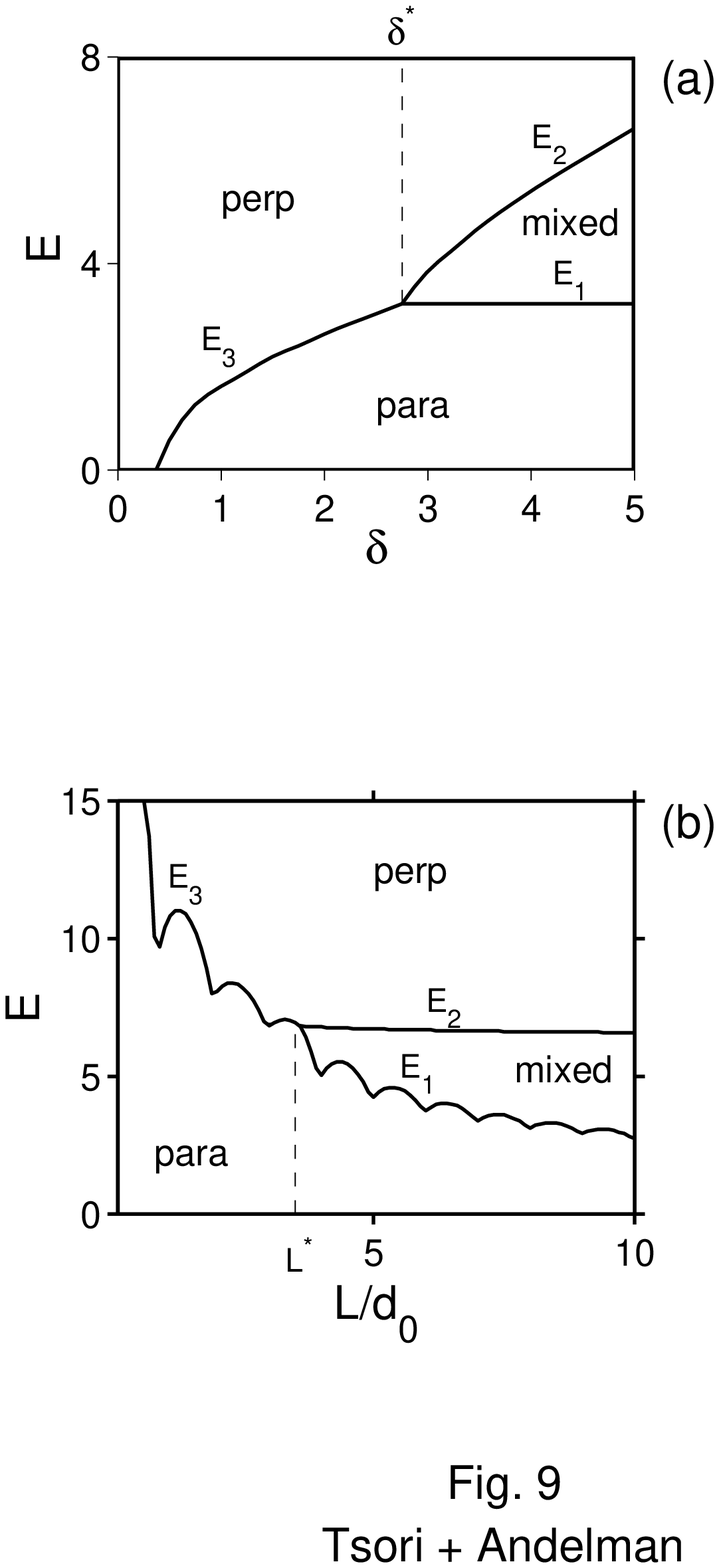}
\end{figure}



\begin{thebibliography}{99}
\bibitem{7:chaikin1} Park, M.; Harrison, C.; Chaikin, P.
M.; Register, R. A.; Adamson, D. H. {\it Science} {\bf
1997}, {\it 276}, 1401.

\bibitem{7:fink1} Ibanescu, M.; Fink, Y.; Fan, S.; Thomas, E. L.;
Joannopoulos, J. D. {\it Science} {\bf 2000}, {\it 289},
415.

\bibitem{7:fink2} Fink, Y.; Winn, J. N.; Fan, S.; Chen, C.; Michel,
J.;
Joannopoulos, J. D.; Thomas, E. L. {\it Science} {\bf
1998}, {\it 282}, 1679. Urbas, A.; Sharp, R.; Fink, Y.;
Thomas, E. L.; Xenidou, M.; Fetters, L. J. {\it Adv.
Matter.} {\bf 2000}, {\it 12}, 812.

\bibitem{7:Leibler80} Leibler, L. {\it Macromolecules} {\bf 1980},
{\it 13}, 1602.

\bibitem{7:O-K86} Ohta, T; Kawasaki, K. {\it Macromolecules} {\bf
1986}, {\it 19}, 2621.

\bibitem{7:B-F90} Bates, F. S.; Fredrickson, G. H. {\it Annu. Rev.
    Phys. Chem.} {\bf 1990}, {\it 41}, 525.

\bibitem{7:amundson93} Amundson, K.; Helfand, E.; Quan, X.;
Hudson, S. D. {\it Macromolecules} {\bf 1993}, {\it 26},
2698.

\bibitem{7:amundson94} Amundson, K.; Helfand, E.; Quan, X. {\it
Macromolecules} {\bf 1994}, {\it 27}, 6559.


\bibitem{7:russellSC96} Morkved, T.; Lu, M.; Urbas, A. M.;
Ehrichs, E. E.; Jaeger, H. M.;  Mansky, P.; Russell, T. P.
{\it Science} {\bf 1996}, {\it 273}, 931.

\bibitem{7:TDR-MM00} Thurn-Albrecht, T.; DeRouchey, J.;
Russell, T. P. {\it Macromolecules} {\bf 2000}, {\it 33},
3250.

\bibitem{7:epje01} Tsori Y.; Andelman, D. {\it Eur.
Phys. J. E} {\bf 2001}, {\it 5}, 605.

\bibitem{7:turnerPRL92} Turner, M. S. {\it Phys. Rev. Lett.} {\bf
1992}, {\it 69}, 1788.

\bibitem{7:W-RMM94} Walton, D. G.; Kellogg, G. J.; Mayes, A. M.;
Lambooy, P.; Russell, T. P. {\it Macromolecules} {\bf
1994}, {\it 27}, 6225.

\bibitem{7:F-H87} Fredrickson, G. H.; Helfand, E. {\it J. Chem.
Phys.} {\bf 1987}, {\it 87}, 697.

\bibitem{7:epl01} Tsori, Y; Andelman, D. {\it
Europhys. Lett.} {\bf 2001}, {\it 53}, 722.

\bibitem{7:mm01} Tsori, Y.; Andelman, D. {\it
Macromolecules} {\bf 2001}, {\it 34}, 2719.

\bibitem{7:sh77} Swift, J.; Hohenberg, P. C. {\it Phys.
Rev. A} {\bf 1977}, {\it 15}, 319.

\bibitem{7:cc98} Chen, H.; Chakrabarti, A. {\it  J. Chem. Phys.}
{\bf 1998}, {\it 108}, 6897.

\bibitem{7:fraaije02} For a calculation which does not assume
$\xi\gtrsim L$ refer to: Kyrylyuk, A. V.; Zvelindovsky, A. V.;
Sevink, G. J. A.; Fraaije, J. G. E. M. {\it Macromolecules} {\bf
2002}, {\it 35}, 1473.

\bibitem{7:onuki95} Onuki, A; Fukuda, J. {\it Macromolecules} {\bf
1995}, {\it 28}, 8788.


\bibitem{7:PW-MM99} Pereira, G. G.; Williams, D. R. M. {\it
Macromolecules} {\bf 1999}, {\it 32}, 8115.

\bibitem{7:hh} de Gennes, P. G.; Prost, J. {\it The Physics of
Liquid Crystals}, Oxford University: New York, 1993.
Helfrich, W. {\it Appl. Phys. Lett.} {\bf 1970}, {\it 17},
531. Helfrich, W. {\it J. Chem. Phys.} {\bf 1971}, {\it
55}, 839. Hurault, J. P. {\it J. Chem. Phys.} {\bf 1973},
{\it 59}, 2068.

\bibitem{7:ashok} Ashok, B.; Muthukumar, M;
Russell, T. P. {\it J. Chem. Phys.} {\bf 2001}, {\it 115},
1559.

\bibitem{7:braz}  Brazovskii, S. A. {\it
Sov. Phys. JETP} {\bf 1975}, {\it 41}, 85.

\bibitem{7:fabre} Fabre, P.; Casagrande, C.; Veyssie, M.;
Cabuil, V.; Massart, R. {\it Phys. Rev. Lett.} {\bf 1990},
{\it 64}, 539.

\end{thebibliography}
\end{document}